\documentclass[aps,prb,twocolumn,superscriptaddress,showpacs,amsmath,amssymb]{revtex4-1}

\usepackage{graphicx}
\usepackage{dcolumn}
\usepackage{bm}

\newcommand{\spm}{$s_{\pm}$}

\begin{document}

\title{Interband  coupling and  transport interband scattering in $s_{\pm}$ superconductors }

\author{V. G. Kogan}
\email{kogan@ameslab.gov}
\affiliation{Ames Laboratory, Ames, IA 50011}

\author{R. Prozorov}
\email[Corresponding author: ]{prozorov@ameslab.gov}
\affiliation{Ames Laboratory, Ames, IA 50011}
\affiliation{Department of Physics \& Astronomy, Iowa State University, Ames, IA 50011}

\pacs{74.20.-z, 74.20.Rp}


\begin{abstract}
A two-band model with repulsive interband coupling and  interband {\it transport} (potential) scattering is considered to elucidate their effects on  material properties. In agreement with previous work, we find that the bands order parameters $\Delta_{1,2}$  differ  and the large  is at the band with a smaller normal density of states (DOS), $N_{n2}<N_{n1}$. However, the bands energy gaps, as determined by the energy dependence of the DOS, are equal due to scattering. For each temperature, the gaps turn zero at a certain critical interband scattering rate, i.e. for strong enough scattering the model material becomes gappless. In the gapless state, the DOS at   the band 2  is close to the normal state value, whereas at the band 1 it has a V-shape with non-zero minimum. When  the normal bands DOS' are mismatched, $N_{n1}\ne N_{n2}$,  the critical temperature $T_c$ is suppressed even in the absence of interband scattering, $T_c(N_{n1})$ has a dome-like shape. With increasing interband scattering, the London penetration depth at low temperatures   evolves from  being exponentially flat to the power-law and even to near linear behavior in the gapless state, the latter being easily misinterpreted as  caused by order parameter nodes.
\end{abstract}

\date{\today}
\maketitle

\section{Introduction}

It is by now an accepted view that the interband scattering in two-band $\pm s$ superconductors suppresses the critical temperature, i.e., has a pair-breaking effect, see e.g. Refs.\,\onlinecite{Sung}-\onlinecite{Chubukov}. The interband coupling and interband scattering are of a particular interest because both are thought to play a special role in physics of two-band materials in general\cite{Chow,Mosk1,Schop} and of the  extensive family of Fe-based compounds, in particular.\cite{Mazin} Theoretical description of multiband situation requires multitude of parameters to represent couplings along with   intra- and inter-band scatterings.\cite{Korsh} For this reason, we focus here on  a model with only interband coupling (repulsive, to have $\pm s$ order parameter) and with a nonmagnetic  interband scattering. Although such a model cannot be applied to real materials, it allows one to single out physical consequences of the interband scattering which may help in data interpretation.

\section{Approach}

Our approach is based on the quasiclassical version of the weak-coupling BCS theory for  anisotropic Fermi surfaces and   order
parameters  $\Delta $. \cite{E}  In the absence of magnetic fields we have for the Eilenberger Green's functions $f({\bm k},\omega)$ and $g({\bm
k},\omega)$:

\begin{equation}
0=2\Delta g   -2\omega f+I \,,\qquad
\label{eil1}
1=g^2+ f^2 \,.
\end{equation}
($\hbar=1$).
Here, $\Delta(\bm k)$ is the order parameter, $\bm k$ is the Fermi momentum $\omega=\pi
T(2l+1)$ with an integer $l$ are   Matsubara frequencies.
The scattering term $I$ is given by the integral over the full Fermi
surface:
\begin{equation}
I( {\bm k} )=\int d^2{\bm q}\,\rho({\bm q})\,W({\bm k},{\bm q})\left[g({\bm k})f({\bm
q})-f({\bm k})g({\bm q})\right]  \label{I}
\end{equation}
with $W({\bm k},{\bm q})$ being the  probability of scattering from   $ {\bm q}$
to  $ {\bm k}$.   The  DOS $\rho({\bm q})$  is
normalized: $\int d^2{\bm q}\,\rho({\bm q})=1$.

 We use approximation of the scattering time $\tau$:
\begin{equation}
\int d^2{\bm q}\,\rho({\bm q})\,W({\bm k},{\bm q})\, \Phi({\bm
q})=\langle\Phi\rangle/\tau\,; \label{tau}
\end{equation}
$\langle ...\rangle$  stands for the average over the  Fermi surface.
Clearly, the approximation amounts to the scattering probability
$W=1/\tau$ being constant for any  ${\bm k}$ and ${\bm q}$.
 However, for two well-separated Fermi surface sheets, the probabilities of intra-band
scatterings may differ from each other and from processes involving ${\bm k}$ and
${\bm q}$ from different bands. The effects of the inter-
 and intra-band scattering upon various properties of the system are
different.
 Hence, Eq.\,(\ref{tau}) is replaced with: \cite{KZ}
 \begin{equation}
\int d^2{\bm q_\nu}\,\rho({\bm q_\nu})\,W({\bm k_\mu},{\bm q_\nu})\,
\Phi({\bm q_\nu})=n_\nu\langle \Phi\rangle_\nu/\tau_{\mu\nu}\,.
\end{equation}
Here  $\nu,\mu =1,2$ are  band indices;  $\langle ...\rangle_{\nu}$
denotes averaging  over the $\nu$-band, and $n_\nu=\int d^2{\bm
q}_\nu\,\rho({\bm q}_\nu)=N_\nu/N(0)$ are relative densities of states:
 $n_1+n_2=1$.

We assume  the order parameter  taking constant
values $\Delta_1$ and $\Delta_2$ at each of the two bands.
Writing Eq.\,(\ref{eil1}) for ${\bm k}$ in the first band, we have:
\begin{eqnarray}
&0&=2\Delta_1 g_1 - 2\omega f_1\nonumber\\ &+& \frac{n_1}{\tau _{11}}(g_1 \langle
f \rangle_1 -f_1\langle g\rangle_1)+ \frac{n_2}{\tau _{12}} (g_1 \langle f
\rangle_2 -f_1\langle g \rangle_2) \,.\qquad
\label{eq5}
\end{eqnarray}
In zero field and with ${\bm k}$ independent $\Delta$'s in
each band,  $f,g$ are ${\bm k}$ independent, i.e.,  $\langle f
\rangle_\nu = f_\nu$ and $\langle g \rangle_\nu=g_\nu$:
 \begin{eqnarray}
0=\Delta_1 g_1 - \omega f_1 + n_2(g_1  f_2 -f_1  g_2)/2\tau _{12}\,.
\label{E1}
\end{eqnarray}
The equation for the second band differs from this by replacement $1\leftrightarrow
2$. The fact that $\tau _{11}$ and $\tau_{22}$ do not enter the system
(\ref{E1}) is similar to the case of one-band isotropic material for which
non-magnetic scattering has no effect upon $T_c$ (the Anderson theorem). It
is the inter-band scattering that makes the difference in the two-band case, the
fact stressed already in early work. \cite{Mosk1,Schop} For brevity, we use the notation $\tau_{12}=\tau$ unless  $\tau _{11}, \tau_{22}$ should be explicitly distinguished from $\tau _{12}$.

Eqs.\,(\ref{E1}) are complemented with normalizations,
\begin{equation}
g_\nu^2+f_\nu^2=1\,,\quad \nu=1,2 \,,
\label{norm}
\end{equation}
and by the self-consistency equation for the order parameter:
 \begin{eqnarray}
\Delta( {\bm k})&=&2\pi TN_n \sum_{\omega >0}^{\omega_D} \Big\langle
V({\bm k},{\bm k}^{\prime\,}) f({\bm k}^{\prime},\omega)\Big\rangle_{{\bm k}^{\prime\,}}.
   \label{self-cons}
\end{eqnarray}
Here, $N_n$ is the total density of states at the Fermi level per spin in the normal phase;
   $\omega_D$ is the Debye frequency (or the energy of whatever  ``glue boson").
  Within the weak-coupling  scheme, the coupling potential  $V$    responsible for
superconductivity is  a $2\times 2$ matrix of constants $V_{\nu\mu}$.
The  self-consistency Eq.\,(\ref{self-cons}) takes the form:\cite{g-model}
\begin{eqnarray}
\Delta_\nu= 2\pi T\sum_{ \mu,\omega}^{\omega_D}  n_\mu \lambda_{\nu\mu} f_\mu \,,\quad \nu=1,2,
   \label{self-cons1}
\end{eqnarray}
 $\lambda_{\nu\mu} = N_nV_{\nu \mu}$ are dimensionless
 coupling constants.

  To separate effects of  the interband coupling and scattering   from other possible multiband consequences, we set $\lambda_{11}=\lambda_{22}=0$, whereas $\lambda_{12}$ (denoted as $\lambda $ in the text below) is assumed {\it negative}. This leads to the order parameters  $\Delta_1$ and $\Delta_2$ having opposite signs, \cite{Geilikman,Mosk1,Mazin} i.e. to $\pm s$ superconductivity, which presumably exists in many Fe-based materials. Hence, we have
  \begin{eqnarray}
 \Delta_1 = 2\pi T\lambda   n_2\sum_\omega^{\omega_D}   f_2 \,,\quad
 \Delta_2 = 2\pi T\lambda   n_1\sum_\omega^{\omega_D}   f_1 \,.
 \label{s-c}
 \end{eqnarray}
 Hereafter, we take $\Delta_1$ as being positive. Since $\lambda<0$, these equations imply negative $\Delta_2$.  Accordingly, in the currents free phase  $f_1>0$ and $f_2<0$; in particular, this prescribes the sign of the square root if the  normalization (\ref{norm}) is used to express $f$'s: $f_1=\sqrt{1-g_1^2}$, $f_2=-\sqrt{1-g_2^2}$.

As in original work by Eilenberger,\cite{E}   the  energy functional $\Omega$ can be constructed so that Eqs.\,(\ref{E1}) and (\ref{s-c})
follow as extremum conditions relative to variations of $f_\nu$ and $\Delta_\nu$:
\begin{eqnarray}
\frac{ \Omega}{N(0)} = \frac{2\Delta_1\Delta_2}{\lambda} - 2\pi T  \sum_{\omega }\Big\{\sum_\nu 2n_\nu[\Delta_\nu f_\nu +\omega(g_\nu-1)]\nonumber\\  +\frac{n_1n_2}{\tau}(f_1f_2+g_1g_2-1)\Big\}  \,.\qquad
 \label{functional}
 \end{eqnarray}
Here, $g_\nu$ are abbreviations for $\sqrt{1-f_\nu^2}$, and $\delta g_\nu=-(f_\nu/g_\nu)\,\delta f_\nu$. If $f_\nu$ are solutions of  Eqs.\,(\ref{E1}) and $\Delta_\nu$ satisfy the self-consistency Eqs.\,(\ref{s-c}), $\Omega$ coincides with the condensation energy $F_S-F_N$ and can be used to study thermodynamic properties of a uniform two-band system.

Equations (\ref{E1}), (\ref{s-c}), and (\ref{functional}) form the basis of our approach. Only in a few simple situations, the results can be obtained in a closed form. In most cases,   the analytic approach, if at all possible, is too cumbersome, and we resort to numerical solutions which are relatively straightforward with available tools, such as Mathematica or Math-Lab.

\section{Clean case}

It is instructive  to begin   with the clean limit, $\tau\to\infty$, although it has been considered in literature.\cite{Moskalenko,Kresin,Bang} In this case, we have from Eqs.\,(\ref{E1}) and (\ref{norm}):
\begin{eqnarray}
f_\nu=\Delta_\nu/\beta_\nu,\quad g_\nu=\omega/\beta_\nu,\quad  \beta_\nu^2=\omega^2+\Delta _\nu^2.
\label{f,g}
 \end{eqnarray}
  At $T=0$, the sums in Eqs.\,(\ref{s-c}) are evaluated replacing $2\pi T\sum_\omega\to \int_0^{\omega_D}d\omega$ that gives:
 \begin{eqnarray}
 \Delta_1 =  \lambda   n_2\Delta_2\ln\frac{2\omega_D}{|\Delta_2|} \,,\quad
 \Delta_2 = \lambda   n_1\Delta_1\ln\frac{2\omega_D}{|\Delta_1|}  \,.
 \label{s-c0}
 \end{eqnarray}
Expressing  the log-factors and subtracting the results, one obtains   for the ratio $R =|\Delta_2/\Delta_1|$:
 \begin{eqnarray}
|\lambda|  \ln R = \frac{R}{n_1}- \frac{1}{n_2R}  \,.
 \label{R0}
 \end{eqnarray}
Given $n_{1,2}$ and $\lambda$, this can be solved numerically for $R$. E.g., for $\lambda=-0.6$ and $n_1=0.6$, $n_2=0.4$,    we obtain $R\approx 1.27$, whereas for $n_1=0.4$, $n_2=0.6$ we have $R\approx 0.79$. Hence, the order parameter value is larger at the band with a smaller DOS.\cite{remark1}

For a given $R$, Eqs.\,(\ref{s-c0}) yield:
\begin{eqnarray}
| \Delta_1| &=&  2\omega_D\exp\left(-\frac{R}{n_1|\lambda|}\right) ,\nonumber\\
|\Delta_2| &=& 2\omega_D \exp\left(-\frac{1}{n_2|\lambda| R}\right).
 \label{D's}
 \end{eqnarray}
 Hence, $n_1|\lambda|/R$ and $n_2|\lambda| R$ are effective coupling constants for the first and second bands, respectively.

To evaluate the condensation energy at $T=0$, consider the sum   which enters the energy (\ref{functional}):
\begin{eqnarray}
4\pi T\sum_{\omega>0}^{\omega_D} n_1\left[\frac{\Delta_1^2}{\beta_1}+\omega\left(\frac{\omega }{\beta_1} -1\right)\right] =2n_1\int_0^{\omega_D} (\beta_1-\omega)d\omega \nonumber\\
= n_1\left(\frac{\Delta_1^2}{2} +\Delta_1^2\ln\frac{2\omega_D}{|\Delta_1|} \right) =  \frac{n_1\Delta_1^2}{2}+\frac{\Delta_1\Delta_2}{\lambda}, \qquad
 \label{eq8}
 \end{eqnarray}
 where Eqs.\,(\ref{s-c0}) have been used. Hence, we  obtain:
\begin{eqnarray}
\left(F_S-F_N\right)_{T=0}= -N_n\frac{n_1\Delta_1^2+n_2\Delta_2^2}{2} =-N_n\frac{\langle\Delta ^2\rangle}{2}. \qquad
 \label{energy0}
 \end{eqnarray}
Recall:   in isotropic one-band superconductors this energy is $-N_n \Delta ^2/2$.

As $T\to T_{c0}(n_1)$ (the critical temperature of a clean material for a given  $n_1$) $f_\nu=\Delta_\nu/\omega$ and the sums in Eqs.\,(\ref{s-c})
 can be evaluated:
 \begin{eqnarray}
\frac{ \Delta_1}{ \Delta_2} =  \lambda   n_2\ln\frac{2\omega_De^\gamma}{\pi T_{c0}} \,,\quad
\frac{ \Delta_2} { \Delta_1}=  \lambda   n_1\ln\frac{2\omega_De^\gamma}{\pi T_{c0}} \,.
 \label{s-c-Tc}
 \end{eqnarray}
Multiplying these, one extracts the log-factor and the critical temperature:
 \begin{eqnarray}
 \pi e^{-\gamma}T_{c0}  =   2\omega_D\exp(- 1/ \tilde{\lambda}  ),\quad   \tilde{\lambda}=|\lambda |\sqrt{n_1n_2} \,.
 \label{Tc0}
 \end{eqnarray}
Hence,
 \begin{eqnarray}
 \tilde\lambda= |\lambda|\sqrt{n_1n_2}
 \label{tilde-lam}
 \end{eqnarray}
 plays the role of the  overall coupling constant.

It is worth noting that for a fixed coupling $\lambda$, the critical temperature $T_{c0}$ as a function of relative DOS $n_1$ has a dome-like shape; see Fig.\ref{f1}.
Thus within the model of exclusively interband coupling, a mismatch of bands DOS' suppresses $T_c$ even in the absence of scattering. Qualitatively, this happens because for $n_1\ne n_2$, the number of unpaired carriers is proportional to  $|n_1- n_2|$.

Turning back to Eqs.\,(\ref{s-c-Tc}) one finds the ratio\cite{Bang}
 \begin{eqnarray}
R(T_{c0})= \Big |\frac{ \Delta_2} { \Delta_1}\Big |  = \sqrt{\frac  {n_1}{n_2}}   \,.
 \label{R(Tc)}
 \end{eqnarray}
Compare this with Eq.\,(\ref{R0}) for $T=0$  to see that in fact $\Delta_2/\Delta_1$ depends on $T$.

Next, we calculate $  \Delta_\nu $ with the help of the self-consistency system (\ref{s-c}). Note first that near $T_{c0}$, $\beta_\nu\approx \omega (1+\Delta_\nu^2/2\omega^2)$ and, therefore,
 \begin{eqnarray}
f_\nu= \frac{ \Delta_\nu} { \omega}-  \frac{ \Delta_\nu^3} { 2\omega^3} +{\cal O}(\delta t)^{5/2}\,,
 \label{f(Tc)}
 \end{eqnarray}
 here $\Delta\propto(\delta t)^{1/2}$,  $\delta t=1-T/T_{c0}$.
The sums in Eq.\,(\ref{s-c}) are:
 \begin{eqnarray}
\sum_0^{\omega_D} \frac{ 2\pi T} { \omega}= \ln\frac{2\omega_De^\gamma}{\pi T_{c0}}=\frac{1}{\tilde\lambda}+\delta t,\,\,\,\,
\sum_0^{\infty} \frac{ 2\pi T} { \omega^3}=\frac{7\zeta(3)}{4\pi^2T_{c0}^2}\qquad
 \label{f(Tc)}
 \end{eqnarray}
 and we obtain:
 \begin{eqnarray}
\Delta_1= \lambda n_2\Delta_2\left( \frac{ 1} { \tilde\lambda}+\delta t-  \frac{7\zeta(3)  } { 8\pi^2T_{c0}^2} \Delta_2^2\right)\,,\nonumber\\
\Delta_2= \lambda n_1\Delta_1\left( \frac{ 1} { \tilde\lambda}+\delta t-  \frac{7\zeta(3)  } { 8\pi^2T_{c0}^2} \Delta_1^2\right)\,.
 \label{s-cTc)}
 \end{eqnarray}
 This system should be solved keeping terms of the order not higher than $(\delta t)^{3/2}$. One substitutes $\Delta_2$ from the second equation to the first to obtain:\cite{Bang}
 \begin{eqnarray}
\Delta_1^2=    \frac{ 16\pi^2T_{c0}^2\delta t }{7\zeta(3)  }  n_2 \,,\quad
\Delta_2^2=    \frac{ 16\pi^2T_{c0}^2\delta t }{7\zeta(3)  }   n_1\,.
 \label{DsTc0)}
\end{eqnarray}
Thus, the gaps ratio  {\it near} $T_{c0}$ is the same as at $T_{c0}$.\cite{Schm}

The energy near $T_{c0}$ should be evaluated including terms of the order   $(\delta t)^2$. In particular,
 \begin{eqnarray}
g_\nu=1- \frac{f_\nu^2} { 2}-  \frac{f_\nu^4} {8} =1- \frac{ \Delta_\nu^2} { 2\omega^2} +\frac{3 \Delta_\nu^4} { 8\omega^4}  \,.
 \label{g(Tc)}
 \end{eqnarray}
 A straightforward algebra shows that  terms of the order $  \Delta ^2\sim \delta t$ cancel out and the rest give:
 \begin{eqnarray}
F_S-F_N= -N(0) n_1n_2  \frac{16\pi^2T_{c0}^2 }{7\zeta(3)  } \left(1-\frac{T}{T_{c0}}\right)^2.
 \label{energyTc}
 \end{eqnarray}
 Thus, the specific heat jump is:\cite{Mosk2,g-model}
  \begin{eqnarray}
\frac{ C_S-C_N}{C_N}\Big|_{T_c}=    \frac{48}{7\zeta(3)  }n_1n_2 \,.
 \label{DC/Cn}
 \end{eqnarray}
The maximum value of this ratio $ 12/7\zeta(3)= 1.43$ is achieved if $n_1=n_2=1/2$.

\section{Effects of scattering}

In general, in the presence of the interband scattering, the system of Eqs.\,(\ref{E1}) and (\ref{s-c}) can be solved only numerically. Near $T_c$ however Eqs.\,(\ref{E1}) can be linearized and $f_\nu$ are readily expressed in terms of $\Delta_\nu$:
\begin{eqnarray}
f_\nu &=& \frac{\Delta_\nu}{\omega^\prime}+\frac{\langle\Delta\rangle}{2\omega\omega^\prime\tau}  \,,\nonumber\\
\langle\Delta\rangle &=&n_1\Delta_1+n_2\Delta_2\,,\quad  \omega^\prime=\omega+1/2\tau\, .
\label{f-linear}
\end{eqnarray}
  Substituting this in the self-consistency system (\ref{s-c})   one obtains a system of linear homogeneous equations for $\Delta_{1,2}$, which has non-trivial solutions only if its determinant is zero. This gives an implicit equation for $T_c$:
\begin{eqnarray}
0&=&1-2n_1n_2\lambda B -n_1n_2\lambda^2 A(A+B)\,,\label{det=0}\\
A&=&\ln\frac{\omega_D}{2\pi T_c}-\psi\left(\frac{\rho_c+1}{2}\right)\nonumber\\
&=&\frac{1}{\tilde\lambda}+\ln\frac{T_{c0}}{T_c}-\psi\left(\frac{\rho_c+1}{2}\right)+\psi\left(\frac{ 1}{2}\right),\label{A}\\
 B&=& \psi\left(\frac{\rho_c+1}{2}\right)- \psi\left(\frac{ 1}{2}\right),\quad  \rho_c=\frac{1}{2\pi T_c\tau}.
\label{AB}
\end{eqnarray}

If  $n_1=n_2=1/2$ and $\tilde\lambda=|\lambda|/2$,
Eq.\,(\ref{det=0}) reduces to a quadratic equation for  $ \ln(T_{c0}/T_c) $. This gives
\begin{eqnarray}
 \ln\frac{T_{c0}}{ T_c}=\ \psi\left(\frac{\rho_c+1}{2}\right)- \psi\left(\frac{ 1}{2}\right),
\label{AG}
\end{eqnarray}
which coincides with the equation for $T_c$ suppression by impurities for a d-wave one-band superconductor, or generally, for order parameters with zero Fermi surface averages.\cite{Openov,K2010}  In particular this means that for this case $T_c$ turns zero at a critical value of interband scattering time $\tau=1/\Delta_0(0)$,  one-half of   the Abrikosov-Gor'kov's value for the effect of magnetic impurities upon one-band s-wave isotropic superconductors.\cite{AG,Maki}

\subsubsection{$\bm {T_c(n_1,\tau)}$ }

Consider now how the critical temperature changes with changing   $n_1$ and  the scattering rate $1/\tau$. Solving Eqs.\,(\ref{det=0})--(\ref{AB}), we have to take into account that the clean case $T_{c0}$ depends on $n_1$.
To proceed with numerical calculations in this particular problem, we normalize the  temperature:
  \begin{eqnarray}
t^* =    \frac{T}{T_{c0}(0.5)}\,,\qquad  t_c^*=\frac{T_c}{T_{c0}(0.5)} \,.
 \label{t}
 \end{eqnarray}
Here, $T_c(n_1,\tau)$ is the actual critical temperature  and $T_{c0}(0.5)$  is the maximum possible critical temperature of the clean material reached at $n_1=0.5$.

\begin{figure}[htb]
\includegraphics[width=8.cm]{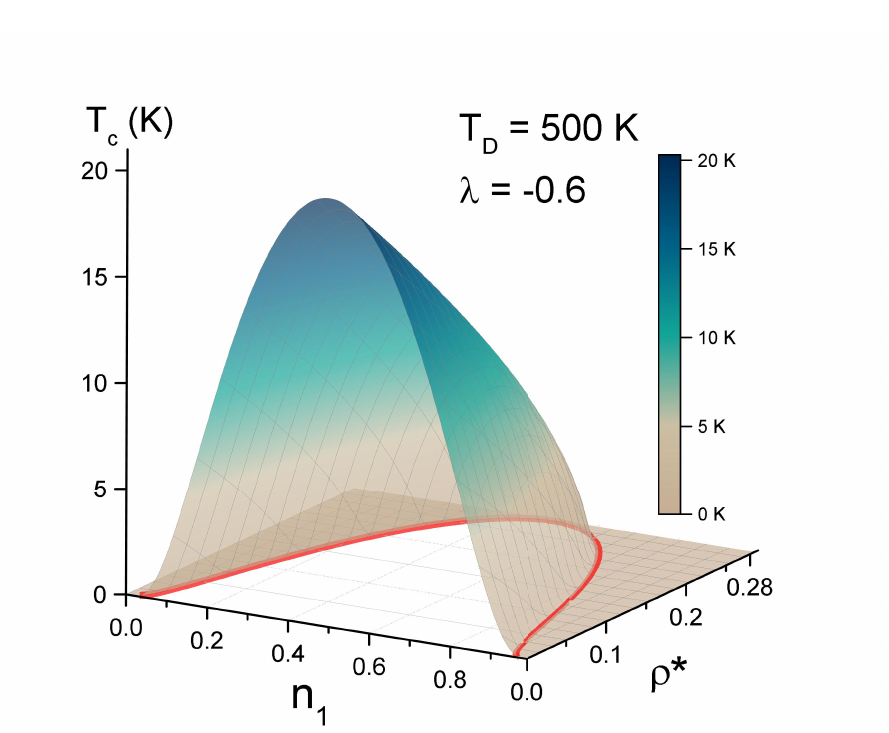}
\caption{(Color online) $T_{c}(n_1,\rho^*)$ for $T_D=500\,$K, $\lambda=-0.6$.  The red line at the dome base  gives the critical value of $\rho^*_{\rm cr}=1/2\pi T_{c0}(0.5)\tau_{\rm cr}$   at which the superconductivity is destroyed. For $n_1=0.5$, $\rho^*_{\rm cr}=e^{-\gamma}/2=0.28$ and the critical rate is $1/\tau_{\rm cr}=\Delta_0(0)$, $\Delta_0(0)$ is the  order parameter of clean material with $n_1=n_2$ at $T=0$.}
\label{f1}
\end{figure}

Also, we introduce the scattering parameters
  \begin{eqnarray}
\rho^*=    \frac{1}{2\pi \tau T_{c0}(0.5)} \,,\quad   \rho_c=    \frac{1}{2\pi \tau T_c}=\frac{\rho^*}{t_c^*}\,,
 \label{r0}
 \end{eqnarray}
 so that   $\rho^*$ is independent  of $n_1$.
 Next, we   transform the log-term in $A$ of Eq.\,(\ref{A}):
\begin{eqnarray}
A =\frac{2}{|\lambda|}-\ln t_c^* +\psi\left(\frac{ 1}{2}\right)-\psi\left(\frac{\rho^*/t_c^*+1}{2}\right)\,.
\label{Aa}
\end{eqnarray}
One should also replace $\rho_c\to \rho^*/t^*_c$ in   $B$ of Eq.\,(\ref{AB}).  The numerical solutions of Eq.\,(\ref{det=0}) for the critical temperature are given in Fig.\,\ref{f1}.

Hence, not only $T_c$ is suppressed by the interband scattering for a fixed $n_1$, but the DOS asymmetry $(n_1-0.5) $ also causes $T_c$  suppression.

One thus concludes that for negative interband coupling $\lambda$, there are two mechanisms for the $T_c$ suppression (pair breaking): the interband {\it transport} scattering and the mismatch of the densities of states of two bands.  In particular, in the presence of interband scattering, the interval of DOS mismatch,  in which the superconductivity exists, shrinks.

\subsubsection{$\bm {T_c(\tau)}$ for a fixed $n_1$}

In the rest of the text, we consider system properties for a fixed normal state DOS $n_1$. It is more convenient to employ reduced temperatures
  \begin{eqnarray}
t  =    \frac{T}{T_{c0}(n_1)}\,,\qquad  t_c =\frac{T_c}{T_{c0}(n_1)}\,.
 \label{t}
 \end{eqnarray}
and the scattering parameters
  \begin{eqnarray}
\rho_0=    \frac{1}{2\pi \tau T_{c0}(n_1)} \,,\quad   \rho_c=    \frac{1}{2\pi \tau T_c}=\frac{\rho_0}{t_c}\,.
 \label{r0}
 \end{eqnarray}

\begin{figure}[htb]
\includegraphics[width=7.cm] {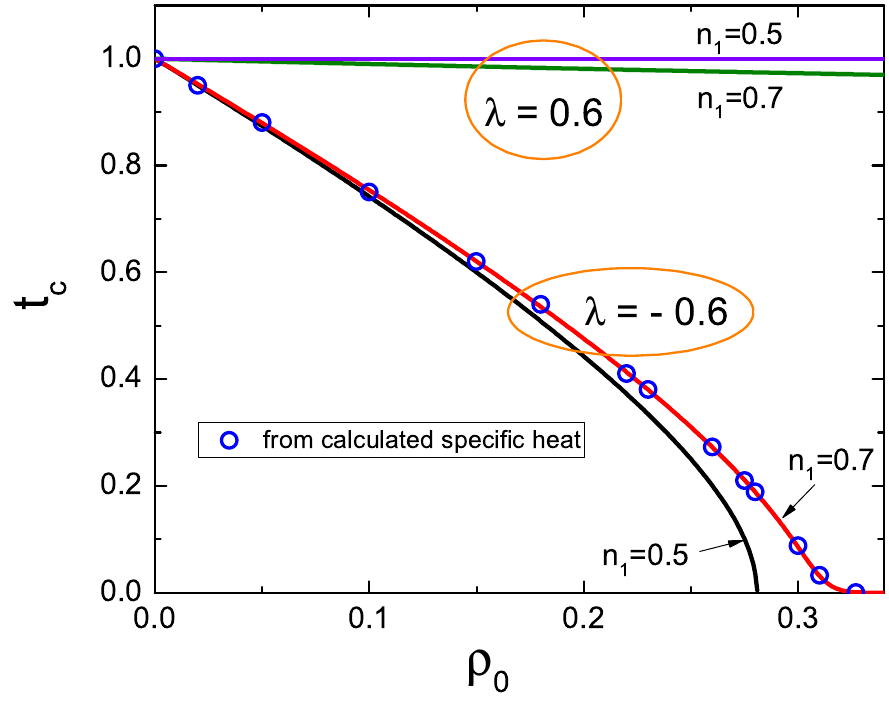}
\caption{(Color online) $t_c=T_c/T_{c0}$ versus $\rho_{0}$ according to Eqs.\,(\ref{det=0})-(\ref{AB}). Lower curves are for $\lambda=-0.6$;  the black line is  for  the  partial DOS $ n_1= 0.5$,     the red line is for $ n_1=0.7$. The upper curves are for positive (attractive) interband coupling constant $\lambda=0.6\,$. The dotes are obtained by independent calculation of the specific heat jumps.}
\label{f2}
\end{figure}

 Fig.\,\ref{f2}   shows  the $T_c(\rho_0)$  for $n_1=0.5$ and 0.7 obtained by solving Eqs.\,(\ref{det=0})-(\ref{AB}). Note that for $n_1=n_2$, the critical value of  $\rho_0$ is $e^{-\gamma}/2\approx 0.28$. Note also that $\rho_0$ characterizes   the scattering along with  the DOS' mismatch. For this reason, the critical value $\rho_{0,\rm cr}$ for $n_1=0.7$ exceeds  0.28 since $T_{c0}(0.7)<T_{c0}(0.5)$.

If $\lambda_{12}>0$, $T_c$ is only weakly reduced by the interband scattering. This behavior   is qualitatively similar to the   one-band s-wave  materials with anisotropic Fermi surfaces, see e.g. Refs.\,\onlinecite{Mosk2}, \onlinecite{g-model}, \onlinecite{Korsh}.
Note, that the $T_c$ suppression is stronger for larger differences of $n_1$ and $n_2$.

\subsection{Order parameters}

To find $\Delta_\nu(T)$ we have to solve the system of Eqs.\,(\ref{E1}) and (\ref{s-c}). Near $T_c$ one can do this analytically and verify that $\Delta_\nu\propto \sqrt{T_c-T}$.
We, however, resort  to  numerical evaluation for arbitrary temperatures and use the analytical limits to verify the  results. We use dimensionless   variables:
\begin{eqnarray}
 \delta_\nu=\frac{\Delta_\nu}{2\pi T_{c0}}\,,\quad t=\frac{T}{ T_{c0}}\,,\quad \rho_0=\frac{1}{2\pi  T_{c0}\tau }\,.
\label{d-var}
\end{eqnarray}
The first of Eqs.\,(\ref{E1}) for $f_1,f_2$ takes the form:
\begin{eqnarray}
 \delta_1g_1- f_1 t(l+1/2)+\frac{n_2\rho_0}{2}(f_2g_1-f_1g_2)=0\,,
\label{Eil1aa}
\end{eqnarray}
where    $l$ is the  Matsubara integer  and $g_\nu=\sqrt{1-f_\nu^2}$.
 The second  equation  is obtained by replacing $1\leftrightarrow 2$.

\begin{figure}[htb]
\includegraphics[width=7.cm] {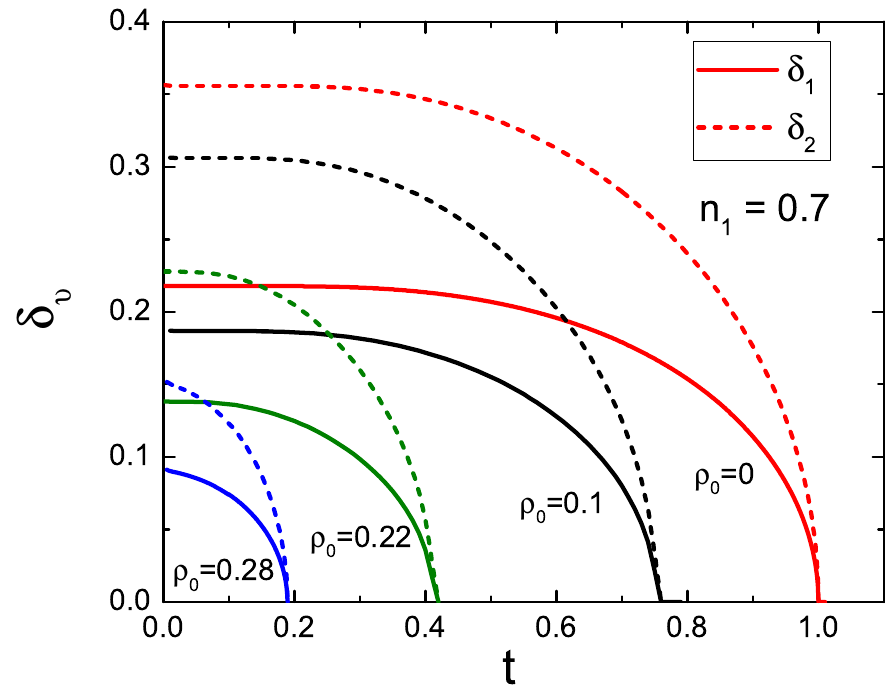}
\caption{(Color online) $|\Delta_\nu|/2\pi T_c $ vs $t=T/T_{c0}$ for $ \lambda=-0.6$, $n_1=0.7$ and   a few values of   $\rho_0=1/2\pi T_{c0}\tau$. }
\label{f3}
\end{figure}

 The first self-consistency Eq.\,(\ref{s-c}) is, see Appendix A:
 \begin{eqnarray}
 \frac{\sqrt{n_1}\delta_1+\sqrt{n_2}\delta_2}{\tilde\lambda \sqrt{n_2}}-\delta_2\ln t =\sum_{i=0}^\infty  \left(\frac{\delta_2}{l+1/2} -t\,f_2\right).\qquad
\label{s-caa}
 \end{eqnarray}
 The second   is obtained by replacing $1\leftrightarrow 2$. Solving the system of four Eqs.\,(\ref{Eil1aa}) and  (\ref{s-caa}) numerically  we obtain $\Delta_\nu(T)$. Examples are shown in Fig.\,\ref{f3}. We note  that, as in the clean case, the order parameter is larger at the band with smaller DOS at all $T$s and for all $\rho_0$.
One sees that near $T_c$, $\Delta_\nu\propto \sqrt{\delta t}$ as it should. This is shown analytically   for $n_1=n_2$ in  Appendix B.

\subsection{Density of states}

\begin{figure}[htb]
\includegraphics[width=7.cm] {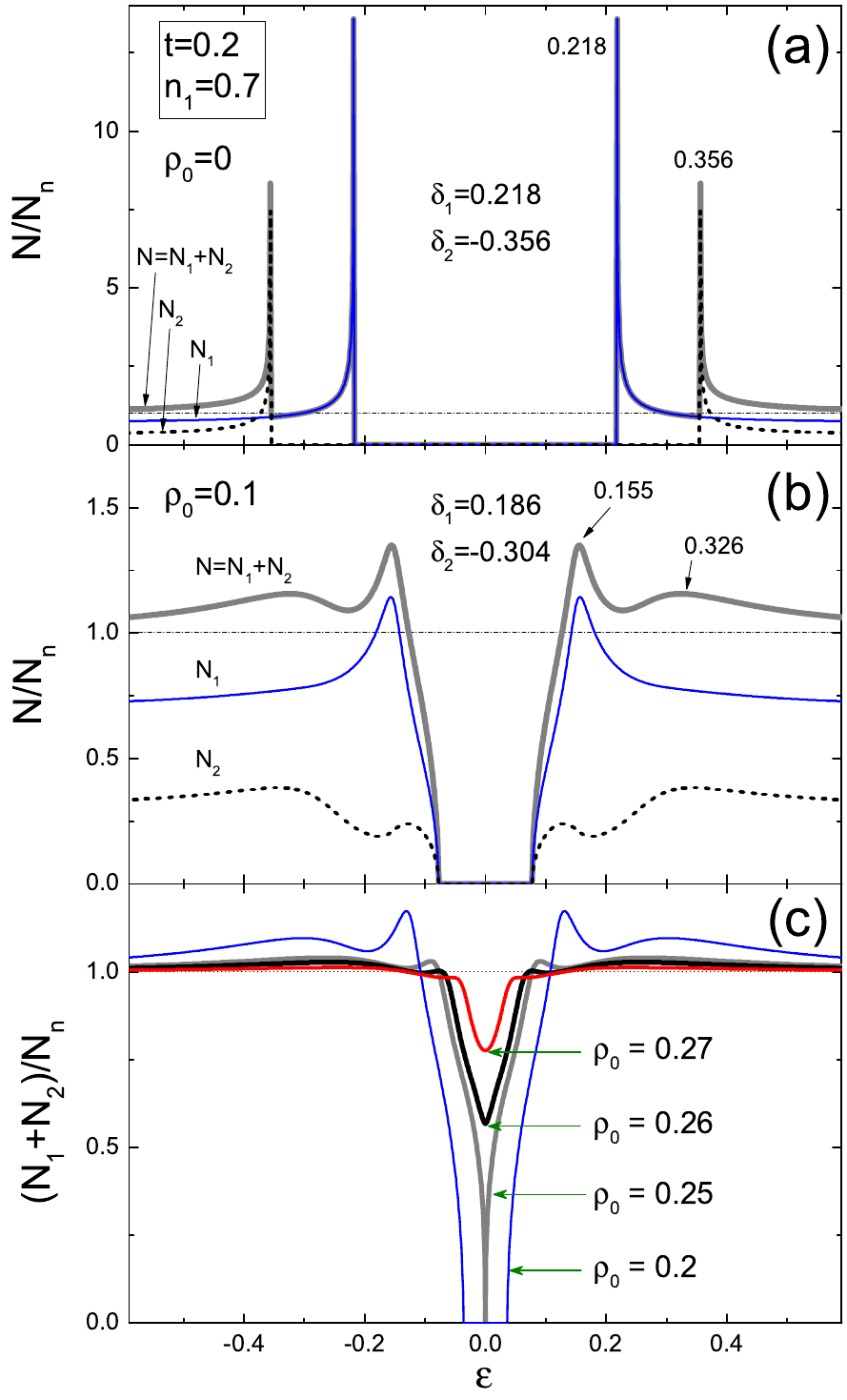}
 \caption{(Color online) (a) The clean limit DOS as a function of energy $\varepsilon=\epsilon/2\pi T_{c0}$ for $ \lambda=-0.6$   and $n_1=0.7$ at $t=0.2$. (b)  The same as (a), but for the interband scattering parameter $\rho_0=0.1$\,.
The bands order parameters for this case  are $\delta_1= 0.186$, $|\delta_2|= 0.304$;  $N(\varepsilon)$ has a typical two-band shape, although the two maxima do not exactly positioned at $|\delta_{1,2}|$. (c) The total DOS for a set of scattering parameters $\rho_0$. Note that with increasing scattering, in the gapless state, the DOS acquires a V-shape with a non-zero minimum.}
 \label{f4}
\end{figure}

As long as $\Delta_\nu(T)$ are known, one can evaluate DOS' $N_\nu$ as  functions of energy $\epsilon$ at any fixed $T$:
\begin{eqnarray}
N_\nu(T,\epsilon) =n_\nu N_n\,{\rm Re}[g_\nu(\omega\to i\epsilon)] \,.
\label{DOS-define}
\end{eqnarray}
To this end, one can  replace  $\omega\to i\epsilon$ already in Eqs.\,(\ref{E1}):
 \begin{eqnarray}
0&=&\Delta_1 g_1 - i\,\epsilon f_1 + n_2\left(g_1  f_2 -f_1  g_2\right)/2\tau  \,, \nonumber\\
f_1 &=&\sqrt{1-g_1^2}\,,\quad f_2 =-\sqrt{1-g_2^2}\,.
\label{E1dos}
\end{eqnarray}
Dimensionless system of equations for $g_\nu$ becomes:
 \begin{eqnarray}
0=\delta_1 g_1 - i\,\varepsilon f_1 + \frac{n_2\rho_0}{2}\left(g_1 f_2  -  g_2 f_1 \right) ,
 \,\,\,\,  \varepsilon=\frac{\epsilon}{2\pi T_{c0}} \qquad
\label{E1dos1}
\end{eqnarray}
(the second equation has $1\leftrightarrow  2$).
The total DOS is $N (T,\epsilon)=N_1(T,\epsilon)+N_2(T,\epsilon)$. Note that DOS depends on $T$ via   $\Delta_\nu (T)$.
Fig.\,\ref{f4} shows  examples of DOS for parameters given in the caption.
The situation is similar to the Abrikosov-Gor'kov  pair-breaking by magnetic impurities where the gap does not coincide with the order parameter.\cite{AG,Maki}
\begin{figure}[htb]
\includegraphics[width=8cm] {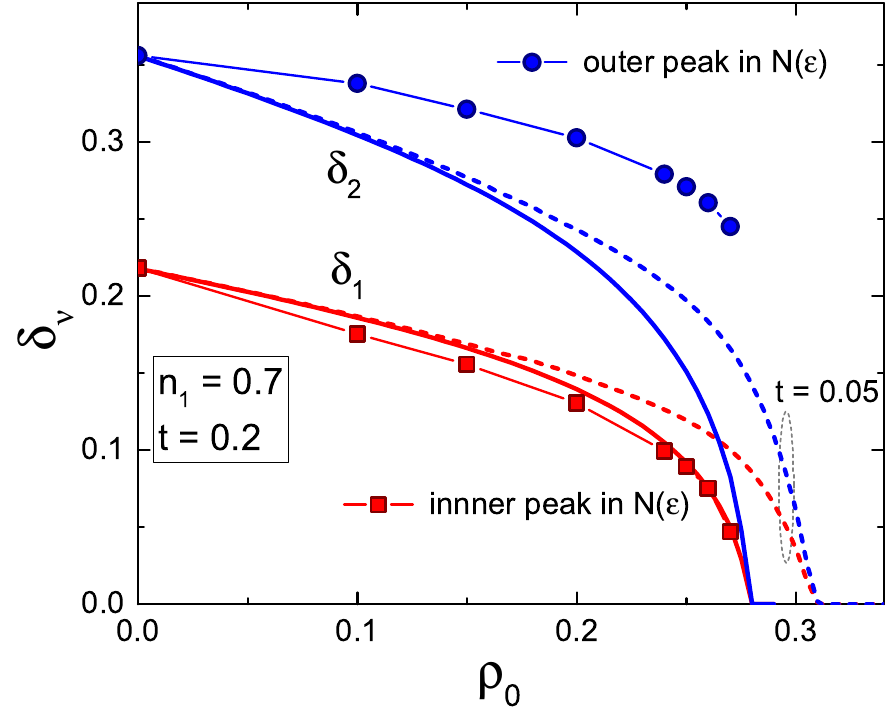}
\caption{(Color online) Peak positions of  DOS $N(\varepsilon)$ vs $\rho_0$ marked as dotes along with the bands order parameters $|\delta_{1,2}|$, solid lines for $t=0.2$. The dashed lines are $|\delta_{1,2}|$ for $t=0.05$. }
\label{f5}
\end{figure}

A remarkable feature of DOS' is worth to note: although $\Delta_1 \ne |\Delta_2|$, the calculated  energy intervals where $N_\nu(\varepsilon)=0$ (the energy gaps) are the same for the two bands, see panel (b) of Fig.\,\ref{f4}. This has  been noticed time ago by Schopohl and Scharnberg who studied two-band model for superconducting transition metals.\cite{Schop}

\begin{figure}[htb]
\includegraphics[width=8cm] {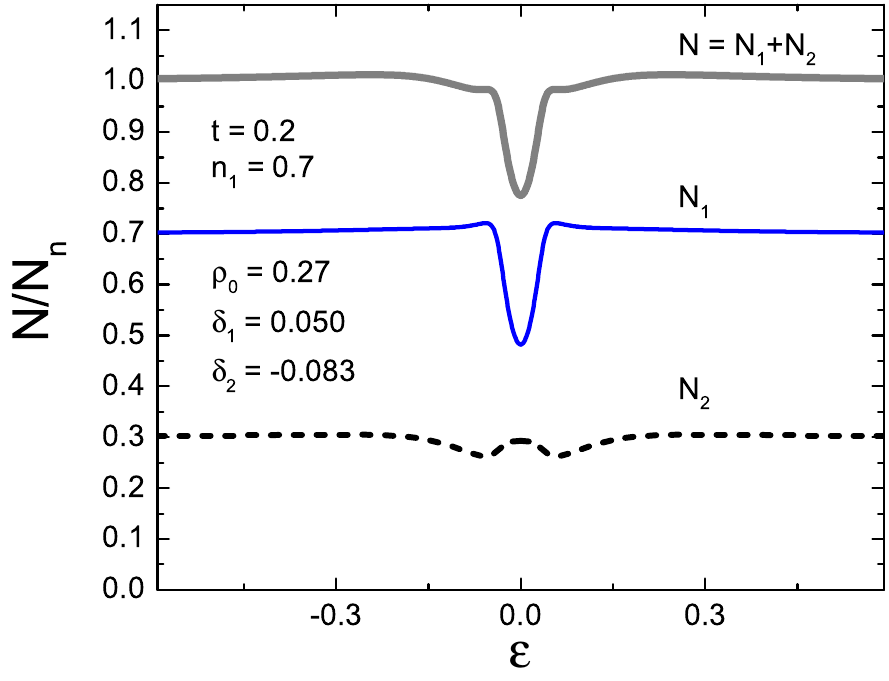}
 \caption{(Color online) The density of states $N$ normalized on $N_n$   vs energy $\varepsilon$ (in units $2\pi T_{c0}$)  for $n_1=0.7$, $t=0.2$  in the gapless state with $\rho_0= 0.27$.}
\label{f6}
\end{figure}

At Fig.\,\ref{f5}b the positions of maxima DOS $N(\varepsilon)$ is shown along with the bands order parameters $|\delta_{1,2}|$ to show that while the first peak is positioned only slightly under $\delta_1$, the second peak is well above  $|\delta_2|$ for all scattering parameters $\rho_0$. This feature has to be taken into account when, e.g., STM data on $N(\varepsilon)$ are interpreted.

It is worth noting that   the energy dependence of DOS $N(\varepsilon)$ in the gapless state, shown in the panel (c) of Fig.\,\ref{f4}, has a ``V" shape which should not  be confused with a similar shape, e.g., in one-band d-wave materials. Another feature worthy of notion is that in the gapless state (in this case $\rho_0>0.25$) the two-band signature is hardly seen. This feature is pronounced in Fig.\,\ref{f6} where both $N_1$ and $N_2$ are shown    for $n_1=0.7$.
We also observe that the band with   $n_2=0.3$  and a larger value of the order parameter ($|\delta_2|=0.083$) has  nearly constant density of states $N_2(\varepsilon)/N_n\approx 0.3 $ at all energies,  close to the normal state value, the fact with implications for, e.g., thermal conductivity.

\subsubsection{Zero-bias DOS ${\bm N_0}$}

At zero energy, the system (\ref{E1dos1}) is simplified. Multiply the first equation by $n_1$, the second by $n_2$ and add them up: $0=n_1\delta_1 g_1+n_2\delta_2 g_2$.
Next,  substitute $g_2=-(n_1\delta_1/n_2\delta_2)g_1$   back to  the first of  Eqs.\,(\ref{E1dos1}) to obtain for $g_1$:
  \begin{eqnarray}
\frac{2\delta_1}{n_2\rho}= \sqrt{1- \frac{n_1^2\delta_1^2}{n_2^2\delta_2^2}g_1^2}-\frac{n_1 \delta_1}{n_2 \delta_2} \sqrt{1-g_1^2 } .
\label{g1}
\end{eqnarray}
 This equation can be resolved relative to $g_1$. After simple algebra one obtains the total zero-energy DOS $N_0$:
  \begin{eqnarray}
\frac{N_0 }{N_n}=\frac{n_1(\delta_2-\delta_1)}{\delta_2} {\rm Re} \sqrt{1- \frac{[(n_2^2\delta_2^2-n_1^2\delta_1^2)\rho_0^2-4\delta_1^2\delta_2^2]^2}{16 n_1^2\rho_0^2\delta_1^4\delta_2^2} }. \nonumber\\
\label{dos}
\end{eqnarray}
\begin{figure}[t]
\includegraphics[width=8cm] {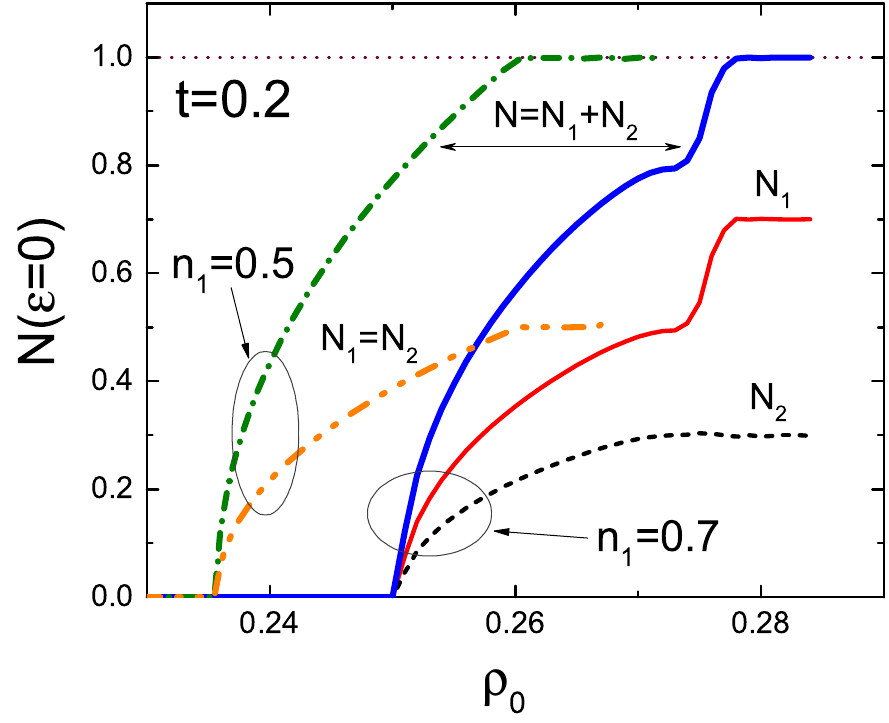}
\caption{(Color online) At the left: the zero-bias DOS (normalizeed to $N_n$) as a function of $\rho_0$ for $ \lambda=-0.6$, $t=0.2$, and   $n_1=0.5$; in this case, $\tilde\rho \approx 0.236$ and $\rho_c\approx 0.26$ so that for $0.236<\rho_0<0.26$ the superconductivity is gapless. At the right: DOS$(\rho_0)$ at zero energy for the same $ \lambda $  and $t$, but  $n_1=0.7$.
}
\label{f7}
\end{figure}
For $n_1=n_2$, $\delta_1=-\delta_2=\delta$, this reduces to
\begin{eqnarray}
\frac{N_0 (\rho_0,T)}{N_n}=  {\rm Re} \sqrt{1- \frac{ 4\delta ^2(\rho_0,T)}{ \rho_0^2} }\,.
\label{dos,n1=n2}
\end{eqnarray}
Clearly,  the solution  of
$\tilde\rho  =2 |\delta(\tilde\rho )| $
  separates the domain $\rho_0<\tilde\rho $ where   $N_0=0$ and the superconductivity is gapped, and the gapless region $\tilde\rho <\rho_0<\rho_c$.

  An example of numerically evaluated DOS for $n_1=0.5$ at $t=T/T_{c0}=0.2$ is the left curve   of Fig.\,\ref{f7}.
The lower boundary of the gapless domain, $\tilde\rho \approx 0.236$,  is $\approx 0.91$ of the critical value 0.26, close to the   estimate for this domain at $T=0$ for magnetic impurities of a single band isotropic material.\cite{AG}

Similarly one can extract an equation for $\tilde\rho $  from Eq.\,(\ref{dos}) for $n_1\ne n_2$:
   \begin{eqnarray}
\tilde\rho  =\frac{2 \delta_1|\delta_2|}{n_1\delta_1+n_2|\delta_2|} \,.
\label{rho*1}
\end{eqnarray}

\begin{figure}[htb]
\includegraphics[width=8cm] {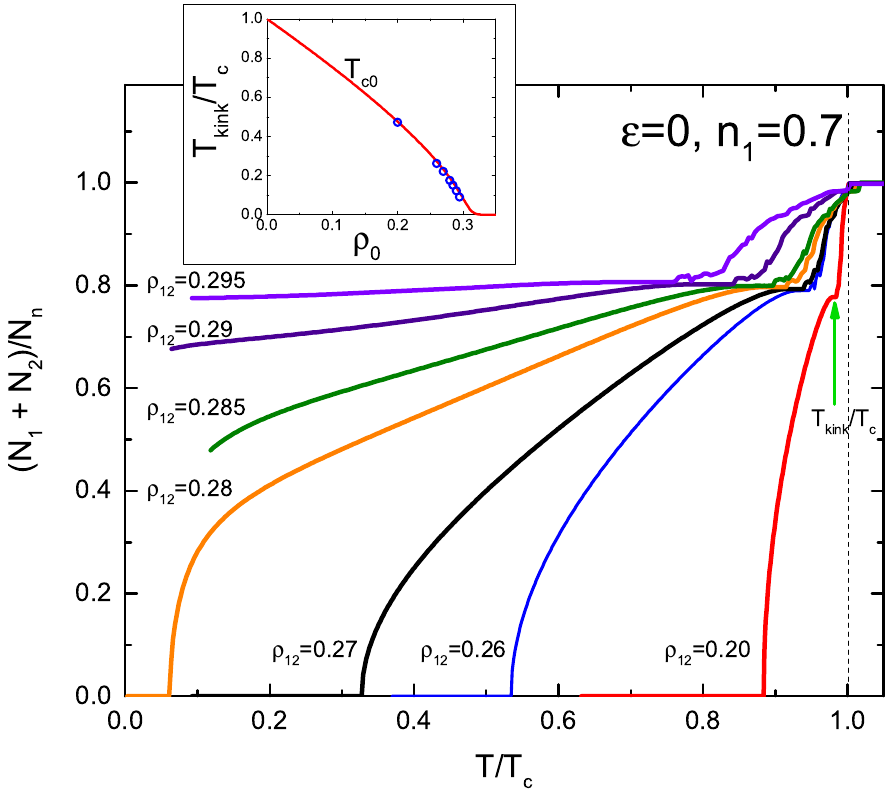}
\caption{(Color online) DOS $N_0/N_n$ at zero energy vs reduced temperature $T/T_c$  for $n_1=0.7$ and a set of scattering parameters indicated.  Note that the temperature is normalized here on actual $T_c$, unlike the most of the text where $T/T_{c0}$ is used.
}
\label{f8}
\end{figure}

An interesting feature of $N_0(\varepsilon)$ seen at the right of Fig.\,\ref{f7}) is a sharp drop near $\rho_0=0.28$ at which $t=0.2$ corresponds to the critical temperature. This feature is seen better yet on the plot of $N$ as a function of temperature at  fixed $\rho_0$ in Fig.\,\ref{f8}.
We observe that the temperature interval of the gapless state near $T_c$ increases with growing $\rho_0$ and covers all $T$'s when $\rho_0\to\tilde\rho$, with $\tilde\rho$ of this case  slightly larger than 0.28. Another feature worth noting is a fast drop of zero-bias $N_0$ near $T_c$, the nature of which at this stage is not clear.

\subsection{Energy and specific heat}

%
\begin{figure}[htb]
\includegraphics[width=8cm] {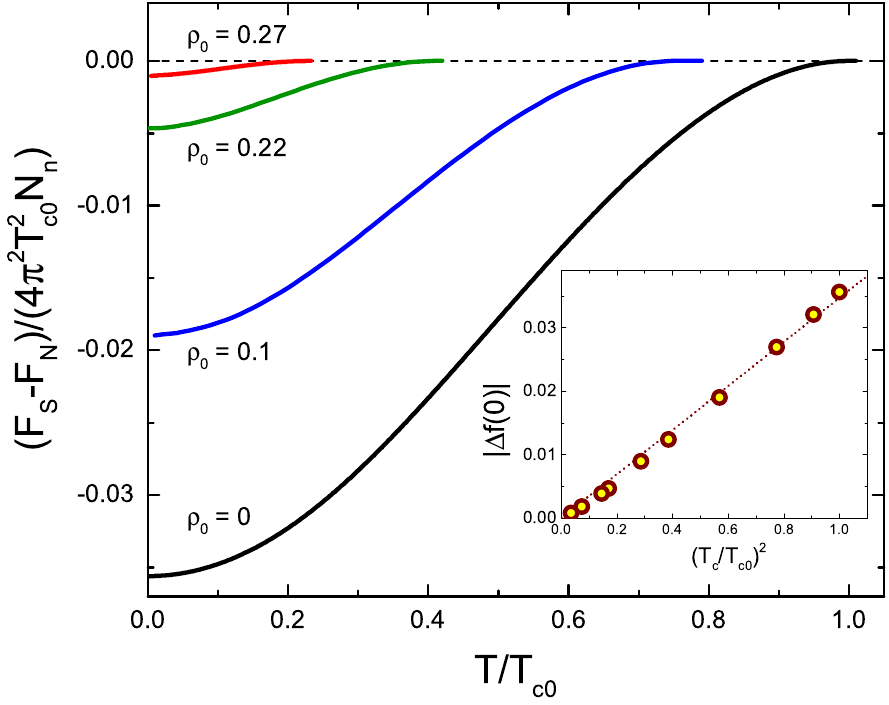}
\caption{(Color online) The temperature dependence of the condensation energy normalized on $4\pi^2T_{c0}^2N_n$ for $n_1=0.7$ and a set of scattering parameters $\rho_0$.  The inset shows that the normalized   condensation energy at $T=0$ scales approximately as $T_c^2$. }
\label{f9}
\end{figure}
\begin{figure}[htb]
\includegraphics[width=8cm] {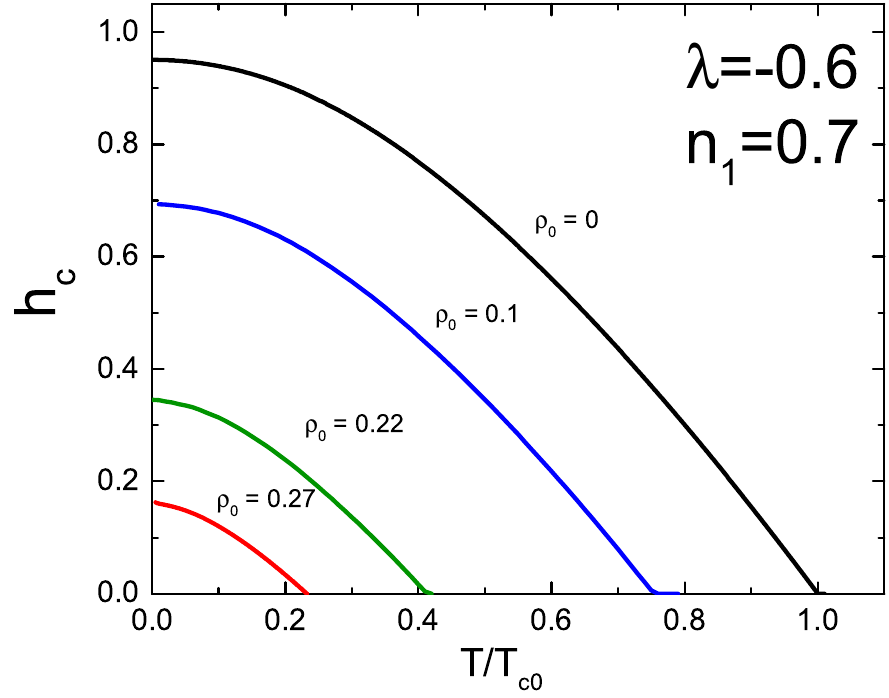}
\caption{(Color online) The thermodynamic critical field $h_c(t)=H_c/H_{c0}$ for $n_1=0.7$ and $ \lambda=-0.6$.}
\label{f10}
\end{figure}

Substituting the self-consistency Eqs.\,(\ref{s-c}) in the functional (\ref{functional}) one obtains:
\begin{eqnarray}
\frac{ \Omega}{N_n} =  - 2\pi T  \sum_{\nu,\omega }n_\nu\left[\Delta_\nu f_\nu +2\omega(g_\nu-1)\right]\nonumber\\
 - 2\pi T\frac{n_1n_2}{\tau} \sum_{\omega }(f_1f_2+g_1g_2-1)   \,.\qquad
 \label{energy}
 \end{eqnarray}
We normalize $\Omega(T)/N_n$  on $4\pi^2 T_{c0}^2$:
\begin{eqnarray}
\frac{ F_n-F_s}{4\pi^2 T_{c0}^2N_n} =   t \sum_{\nu,l }n_\nu\left[\delta_\nu f_\nu   +t(2 l+1)(g_\nu-1)\right]\nonumber\\
+ t\, n_1n_2\rho_{12}   \sum_l(f_1f_2+g_1g_2-1)   \,.\qquad
 \label{energy3}
 \end{eqnarray}
 Since we can calculate $\delta_\nu$ and $f_\nu$ at a given temperature, it is an easy task to evaluate the condensation energy, see Fig.\,\ref{f9}. The inset to this figure shows that the normalized   condensation energy at $T=0$ scales approximately as $T_c^2$, a nearly universal property of all superconductors.\cite{Carbotte,Stewart}

Having the condensation energy, one finds the thermodynamic critical field $H_c =\sqrt{8\pi(F_N-F_S)}$. We normalize it to the zero-$T$ value $H_c^{(0)} =\sqrt{4\pi N(0)}\Delta_0(0)$ for the clean case and $n_1=n_2$ to get:
\begin{eqnarray}
h_c(t)=\frac{H_c(t)}{H_c^{(0)} }= 2\sqrt{2}\,e^\gamma \sqrt{\Phi (t)}  \,,
\label{jHc}
 \end{eqnarray}
 where $\Phi  (t )$ is the RHS of Eq.\,(\ref{energy3}).
With this normalization, the clean limit $h_c(0)=1$ for $n_1=n_2$.

\begin{figure}[htb]
\includegraphics[width=8cm] {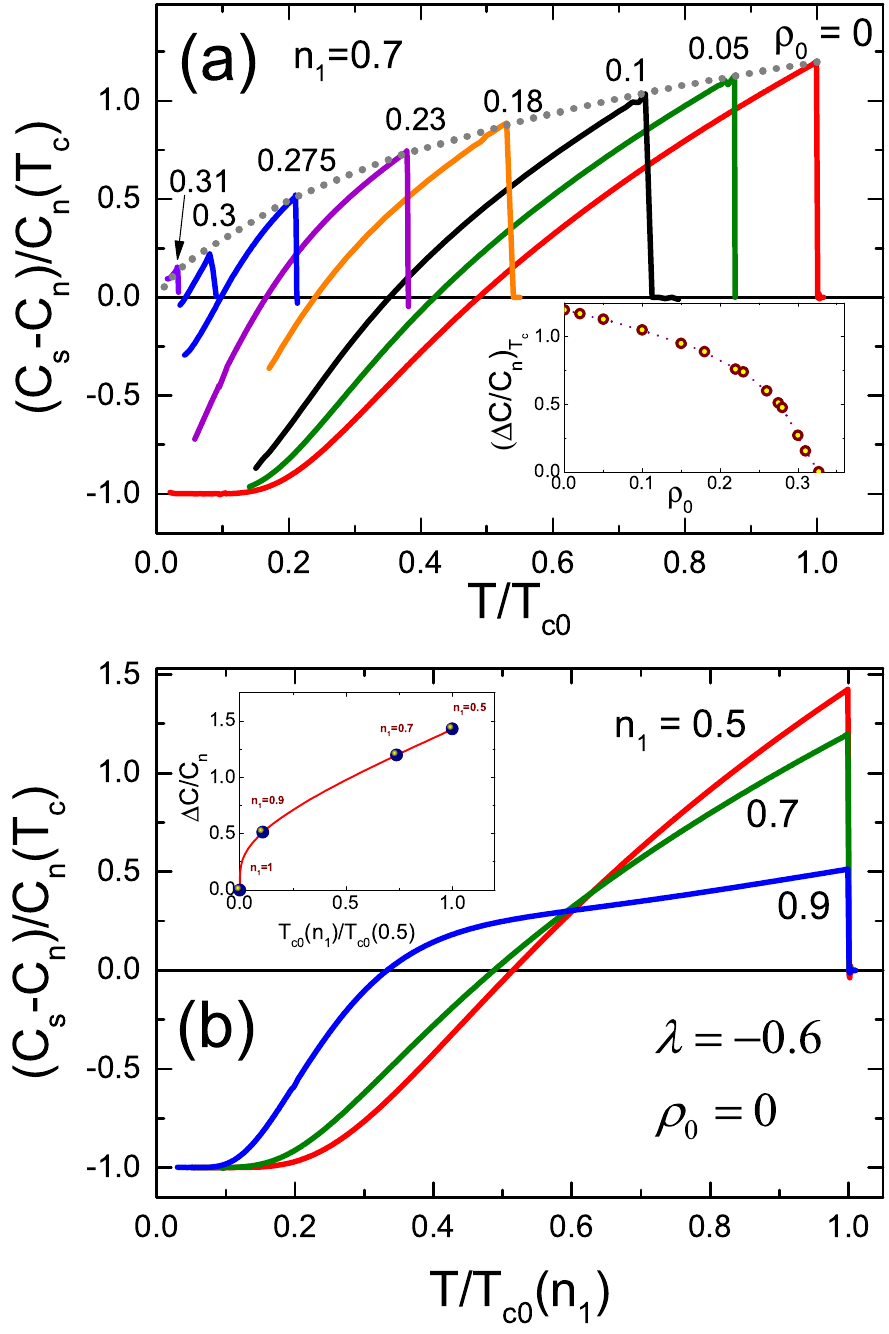}
\caption{(Color online) The upper panel: the specific heat vs $T/T_{c0}(0.7)$ for a few scattering parameters $\rho_0$. The lower panel:  the specific heat vs $T/T_{c0}(n_1)$ for $n_1=0.5,\,\,0.7$ and 0.9. Inset: the specific heat jump at the critical temperature calculated numerically (dots) and according to Eq.\,(\ref{jump(Tc)clean}), the solid line.}
\label{f11}
\end{figure}

The specific heat  can now be evaluated for  fixed $n_1$ and $\rho_0$.
 An example is shown in the upper panel of Fig.\,\ref{f11}.
 The lower panel of Fig.\,\ref{f11} shows the specific heat vs reduced temperature  for a few $n_1$ of clean materials.
Note that the jump at $T_c$ in this case is given in Eq.\,(\ref{DC/Cn}) as a function of  $n_1,\,\,n_2$. On the other hand, $T_{c0}(n_1)$ is given in Eq.\,(\ref{Tc0}) which allows one to evaluate the jump $\Delta C/C_n$ as a function of $T_{c0} $:
\begin{eqnarray}
 \frac{\Delta C}{C_n}\Big |_{T_c} =\frac {48}{7\zeta (3)\lambda^2}\left(\ln\frac{T_{c0}(n_1)}{T_{c0}(0.5)}-\frac{2}{|\lambda|}\right)^{-2}  \,.
 \label{jump(Tc)clean}
  \end{eqnarray}
The inset in the lower panel shows this dependence.
For $n_1=n_2$, analytic evaluation of the specific heat jump is done in Appendix B for any scattering rate.

\subsection{Penetration depth}

If the ground state functions (called
 $f^{(0)}$, $g^{(0)}$ in this section) are known, one can study  perturbations of the
uniform state by a weak magnetic field, i.e., the problem of
the London penetration depth. The perturbations, $f^{(1)},\,\,g^{(1)}$, should be
found from the Eilenberger equations which include gradient terms and magnetic field. \cite{E}  We have for the first band:\cite{KZ}
\begin{eqnarray}
{\bm v_1}{\bm \Pi}f_1 =2\Delta_1 g_1 - 2\omega f_1 + \frac{ n_1}{ \tau _{11}}
[g_1\langle f\rangle_1 -f_1\langle g\rangle_1] \nonumber\\
 + \frac{n_2}{ \tau _{12}}[g_1 \langle f
 \rangle_2 -f_1\langle g \rangle_2] \,,
\label{eq39}
\end{eqnarray}
Here,  ${\bm v}$ is the Fermi velocity, ${\bm \Pi} =\nabla +2\pi i{\bm
A}/\phi_0$ with the vector potential $\bm A$. The second equation is obtained by $1\leftrightarrow 2$.
Two equations for the ``anomalous" functions $f^+$ are obtained from these by
complex conjugation and by ${\bm v}\to -{\bm v}$. \cite{E}  Normalizations
$g_\nu^2+f_\nu f_\nu^+=1$   complete the system.

We now note that  the London approximation  suffices for the problem of weak field penetration. In this approximation    only the overall macroscopic phase $\theta$
depends on coordinates whereas the order parameter modulus remains unperturbed. We thus replace $\Delta \to \Delta e^{i\theta(\bm r)}$ and
look for solutions in the form
\begin{eqnarray}
f_{\nu}&=&(f_{\nu}^{(0)}+f_{\nu}^{(1)})\,e^{i\theta({\bm r})},
\quad  f_{\nu}^+=(f_{\nu}^{(0)}+f_{\nu}^{(1)+})e^{-i\theta({\bm
r})}\,,\nonumber\\
\label{London_approx}
g_{\nu}&=&g_{\nu}^{(0) }+g_{\nu}^{(1) } \,,\qquad \nu=1,2
\,.
\end{eqnarray}
Note that the first  corrections $f_{\nu}^{(1)}), g_{\nu}^{(1) }$ depend on  $\bm k$  (or $\bm v$)   in the form ${\bm v}{\bm
P}$ with ${\bm P}=\nabla\theta+2\pi{\bm A}/\phi_0$, so that their Fermi surface averages vanish.

We obtain  for the corrections in the first band:
\begin{eqnarray}
&&g_1^{(1)}\Delta_1^{\prime}-f_1^{(1)}\omega^{\prime}_1=if_1^{(0)}{\bm v_1}{\bm
P}/2\,,\nonumber\\
&&g_1^{(0)}g_1^{(1)}+f_1^{(0)} f_1^{(1)} =0\,,\label{**}
\end{eqnarray}
where
\begin{eqnarray}
\Delta_1^{\prime}&=&\Delta_1
+n_1 f_1^{(0)}/2\tau_{11} + n_2f_2^{(0)}/2\tau_{12}\,,
\label{D'}\\
\omega^{\prime}_1&=&\omega
+n_1g_1^{(0)}/2\tau_{11}+n_2g_2^{(0)}/2\tau_{12} \,\label{w'}
\end{eqnarray}
contain only the unperturbed  $f^{(0)},g^{(0)}$.
System (\ref{**}) yields:\cite{remark2}
\begin{equation}
g_1^{(1)}=\frac{if_1^{(0)2}\,{\bm v_1}{\bm P}}{2(\Delta_1^{\prime}f_1^{(0)
}+\omega^{\prime}_1g_1^{(0)})}
= i\,\frac{f_1^{(0)2}g_1^{(0)}}{2 \omega^{\prime}_1}\,{\bm v_1}{\bm P}\,.
\label{g^1}
\end{equation}
The correction  $g_2^{(1)}$ for the second band is obtained by replacement $1\to 2$.

\begin{figure}[tbh]
\includegraphics[width=8cm] {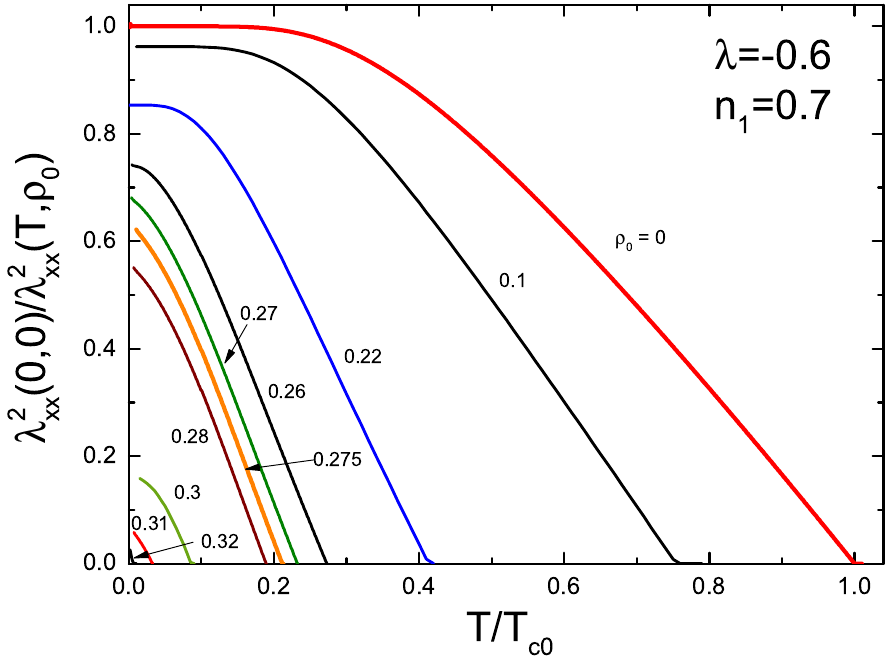}
\caption{(Color online) The inverse square of the in-plane penetration depth normalized on the zero-$T$ clean limit value vs $t=T/T_{c0}$ for   a set of scattering parameters $\rho_0$. In this calculation $\langle v_x^2\rangle_1/\langle v_x^2\rangle_2= 1 $ and the intraband  $\rho_{11}=\rho_{22}=0$.
}
\label{f12}
\end{figure}

To evaluate the penetration depth we turn to the Eilenberger expression for the current density, \cite{E}
\begin{equation}
{\bm j}=-4\pi |e|N_nT\,\, {\rm Im}\sum_{\omega >0}\langle {\bm v}g\rangle\,,
\label{current}
\end{equation}
where $\langle {\bm v}g\rangle=\langle {\bm v}g^{(1)}\rangle $ since  $\langle {\bm v}g^{(0)}\rangle=0$.  Substitute here $g_{\nu}^{(1)}$  of Eq.\,(\ref{g^1}) and
compare with the London relation
\begin{equation}
\frac{4\pi}{c}  j_i=-(\lambda^2)_{ik}^{-1}
\left(\frac{\phi_0}{2\pi}\nabla\theta+{\bm A}\right)_k\,.
\label{London}
\end{equation}
Here, $(\lambda^2)_{ik}^{-1}$ is the tensor of the inverse squared
penetration depth; summation over $k$ is implied.
Hence,    the in-plane component of this tensor is:
\begin{eqnarray}
 \lambda^{-2} _{xx}=\frac{16\pi^2e^2N_nT}{c^2} \sum_{\nu,\omega}
n_\nu\langle
v_x^2\rangle_\nu\frac{f_\nu^2g_\nu}{\omega^\prime_\nu} \,.
\label{lambda}
\end{eqnarray}
Only the unperturbed functions $f^{(0)},g^{(0)}$ enter the penetration depth; for
brevity we dropped  superscripts $(0)$.
 Since we know how to evaluate $f$'s at each temperature, the evaluation of the London penetration depth is straightforward.

For numerical work we normalize $ \lambda^{-2} _{xx}(T,\rho_0)$ on  the zero-$T$ value for  clean bands:
\begin{eqnarray}
 \lambda^{-2} _{xx}(0,0)=\frac{8\pi e^2N_n }{c^2} \langle v_x^2\rangle =\frac{8\pi e^2N_n}{c^2} \sum_{\nu } n_\nu\langle v_x^2\rangle_\nu .\qquad
\label{lambda0}
\end{eqnarray}
Hence, we have for the dimensionless penetration depth:
\begin{eqnarray}
\Lambda^{-2} _{xx} =\frac{\lambda^{-2} _{xx}(T,\rho_0)}{ \lambda^{-2} _{xx}(0,0)}=\frac{  \sum_{\nu,\omega} n_\nu\langle
v_x^2\rangle_\nu f_\nu^2g_\nu /\eta_\nu  } {\sum_{\nu } n_\nu\langle v_x^2\rangle_\nu} \,,\quad\qquad\label{lam-dmless}\\
\eta_\nu = l+\frac{1}{2} + \frac{n_\nu g_\nu\rho_{\nu\nu}}{2t} + \frac{n_{\bar\nu} g_{\bar \nu}\rho_{12}}{2t},\quad
\rho_{\mu\nu}=\frac{\hbar}{2\pi T_{c0}\tau_{\mu\nu}} .\qquad
\label{eta}
\end{eqnarray}
Here, $g_\nu=\sqrt{1-f^2_\nu}$, $\bar\nu$ denotes the value other than $\nu$;   in fact, $\Lambda^{-2}  $ depends only on the ratio of averaged Fermi velocities.

Numerically evaluated   $\Lambda^{-2} _{xx}(t)$ is shown in Fig.\,\ref{f12} for scattering parameters indicated. In this particular calculation  $\rho_{11}=\rho_{22}=0$; incorporating the intraband scattering does not change qualitatively the behavior of the superfluid density with respect to interband scattering and will be presented elsewhere.

\begin{figure}[htb]
\includegraphics[width=8cm] {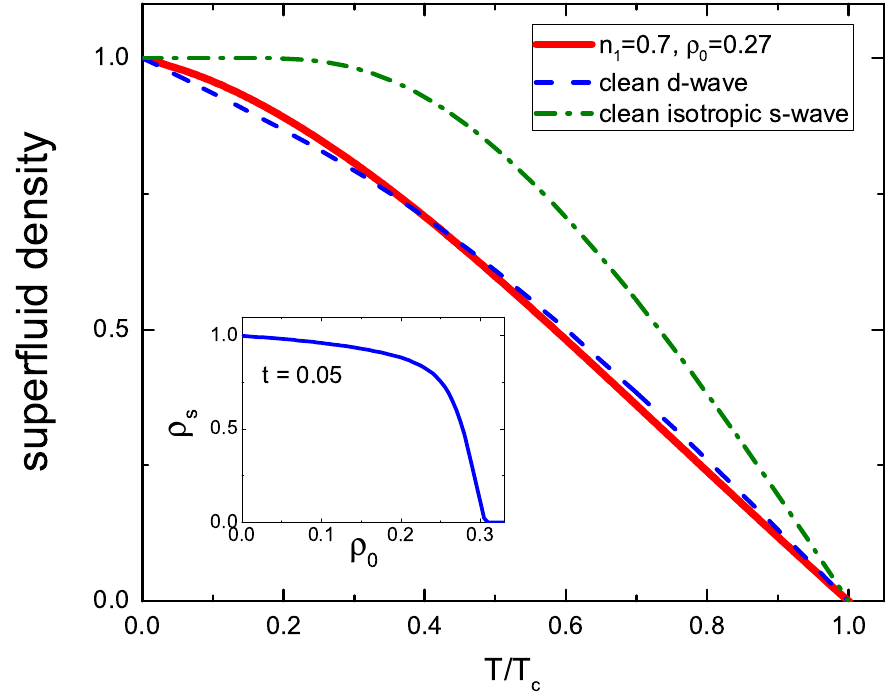}
\caption{(Color online) The superfluid density $\rho_s$  vs $t=T/T_{c0}$ for $n_1=0.7$ and $\rho_0=0.27$ of the gapless state normalized on the value at $T=0$.  Superfluid densities for s- and d-wave clean cases  are shown  for comparison.}
\label{f13}
\end{figure}

We note that for a weak interband scattering the low temperature superfluid density (SFD) is nearly $T$ independent, as expected for  gapped materials. With increasing interband scattering, the flat domain of SFD shrinks and disappears altogether in the  gapless state starting roughly with $\rho_0\approx 0.27$. Remarkably, in the gapless state SFD becomes close to linear, the behavior commonly ascribed to the order parameter nodes. To show that this interpretation can be misleading, we plot SFD for $\rho_0= 0.27$ along with the known result for the d-wave materials in Fig.\,\ref{f13}.

 \section{Discussion}

Many Fe-based compounds are thought to have $\pm s$ symmetry of the order parameter. By considering a model with   the interband coupling $\lambda_{12}<0$ (repulsion) we assure that the bands order parameters $\Delta_1$ and $\Delta_2$ have opposite signs.

Using the quasi-classical Eilenberger approach, we formulate  equations governing two-band systems with  the exclusively interband coupling and  interband scattering. To describe thermodynamic properties  we construct the energy functional,  minimization of which gives the two-band Eilenberger equations along with the   self-consistency equations. This allows us to evaluate the condensation energy along with the specific heat and, in particular, the specific heat jump at $T_c$.

Except some limiting cases which can be dealt with analytically,   we resort to numerical solutions which have  advantage of being straightforward, especially  when analytic approach is too cumbersome if at all possible. For completness we reproduce some of the known results within our approach.

We focus on properties which are affected by the pair-breaking character of the interband scattering. The question of pair-breaking in Fe-based materials has been raised in the past, basically on the basis of Abrikosov-Gor'kov work on magnetic impurities, see, e.g., Refs.\,\onlinecite{K2009}, \onlinecite{K2010}. However, the source of the pair-breaking  was not specified, so that this approach was not generally accepted. Still, formally it seemed to describe a number of observed properties  such as the power-law low temperature dependence of  the superfluid density\cite{Gordon} or the experimentally observed scaling of the specific heat jump $\Delta C\propto T_c^3$.\cite{BNC}

Interband scattering by non-magnetic disorder have qualitatively similar pair-breaking features. In fact, for two bands with equal DOS', the $T_c$ suppression is described by the Abrikosov-Gor'kov Eq.\,(\ref{AG}) for a one-band d-wave material. By evaluating the energy dependence of the density of states, we show that sufficiently strong non-magnetic interband scattering results in a gapless state and  we determine the range of scattering parameters where this state emerges.

The presence of two bands, however, brings in an extra feature: the critical temperature is suppressed not only by the interband scattering but also by a mismatch of bands DOS' $n_1$ and $n_2$. The $T_c$ dependence on $n_1$ has a dome-like shape of Fig.\,\ref{f1}, which suggests that the ubiquitous domes $T_c(x)$ at phase diagrams of, e.g., Fe-based compounds ($x$ is the doping variable) could be related to changing with $x$ of the DOS' mismatch of bands involved. The ability of the model with the interband coupling and scattering to reproduce this dome structure is one of  our main results.

  It is worth noting that the strong pair breaking regime when $T_c\to 0$ in a two-band system with non-magnetic interband scattering differs from the  strong spin-flip scattering by magnetic impurities. The point is that the latter is always complicated by possibility of moments ordering  or by glassy and Kondo phenomena, which are clearly absent for the transport interband scattering.

Properties of the gapless state in the two-band case are richer than in the one-band Abrikosov-Gor'kov situation. Interesting in particular are   properties of DOS in the gapless state. We show that whereas the energy dependence $N_1(\varepsilon)$ of the ``major" band with larger normal state DOS $n_1$  has the ubiquitous V-shape,  the DOS on the ``minor" band is close to being normal.  This suggests a high heat conductance often seen in Fe-based compounds.

Turning to our results on effects of the interband scattering upon the penetration depth, it is instructive to recall the experimental situation. What is commonly
 measured with high accuracy are changes in the London penetration depth, $\Delta \lambda (T) \equiv \lambda(T) - \lambda(0)$. At low temperatures, these are related to the   superfluid density   $\rho_s \equiv \lambda(0)^2/\lambda(T)^2  \approx 1-2\Delta \lambda/\lambda(0)$. It is convenient to analyze low-temperature behavior as $\Delta \lambda (T) \approx 1- AT^n$. According to conventional picture,   the line nodes of the order parameter result in a linear behavior, $n=1$, whereas   fully gapped order parameters (e.g., $s_{++}$ or \spm ) give nearly flat  exponential variation, which in practice is indistinguishable from   $n > 3$.

In the presence of  symmetry-imposed line nodes (e.g., $d$-wave),  intensifying transport scattering causes   monotonic increase of the exponent from $n=1$ to $n=2$,  \cite{Hirschfeld1993,Kogan13pairbreaking} whereas in   the conventional $s$-wave (including multiband $s_{++}$) the low temperature SFD $\rho_s(T)$ remains exponentially flat (whereas $T_c$ does not change).

However, we show in this work that for  fully gapped   \spm \,\, pairing, where potential interband scattering is pair-breaking, the superfluid density  evolves from exponentially flat to nearly linear as shown in Figs.\,\ref{f12} and \ref{f13}. The corresponding exponents in   power-law fits  would change from $n>3$ to   well below $n=2$. In fact, for a strong $T_c$ suppression, in the gapless regime, the entire curve of $\rho_s(T)$ is surprisingly close to a clean $d$-wave dependence, see Fig.13. Thus, in principle, one can change the s-wave-like  to the d-wave-like behavior of $\rho_s(T)$ just by introducing disorder, resulting in a change of the interband scattering. Interesting enough, such a behavior  has been seen in
BaFe$_2$As$_2$ doped with Co or Ni: the exponent $n$ decreased after irradiation.\cite{Kim2010}

\section{Acknowledgment}

This work was supported by the U.S. Department of Energy (DOE), Office of Science, Basic Energy Sciences, Materials Science and Engineering Division. Ames Laboratory is operated for the U.S. DOE by Iowa State University under contract DE-AC02-07CH11358.

\appendix

\section{Self-consistency equations}

Consider the first of self-consistency Eqs.\,(\ref{s-c}):
\begin{eqnarray}
-\frac{\Delta_1}{|\lambda| n_2} =2\pi T\sum_\omega^{\omega_D} f_2 \,.
\label{s-ca}
 \end{eqnarray}
Add and subtract to the RHS $2\pi T\sum_\omega^{\omega_D}(\Delta_2/\omega)$ to have
\begin{eqnarray}
2\pi T\sum_\omega^{\omega_D} \frac{\Delta_2}{\omega} - 2\pi T\sum_\omega^{\infty} \left(\frac{\Delta_2}{\omega} -f_2\right)\,.
\label{s-ca}
 \end{eqnarray}
In the second convergent sum, $\omega_D$ can be replaced with $\infty$, whereas for the first sum  use the identity
\begin{eqnarray}
2\pi T\sum_\omega^{\omega_D} \frac{1}{\omega}  = \frac{1}{\tilde\lambda} -\ln\frac{T}{T_{c0}} \,,\quad \tilde\lambda=| \lambda|\sqrt{n_1n_2}\,.
\label{id}
 \end{eqnarray}
We then obtain:
\begin{eqnarray}
\Delta_2\ln\frac{T_{c0}}{T} &+&\frac{\sqrt{n_1}\Delta_1+\sqrt{n_2}\Delta_2}{\tilde\lambda \sqrt{n_2}}
\nonumber\\
&=& 2\pi T\sum_\omega^{\infty} \left(\frac{\Delta_2}{\omega} -f_2\right)\,.
\label{s-ca}
 \end{eqnarray}

\section{The case $\bm {n_1=n_2}$}

 In this case $\Delta_1=-\Delta_2=\Delta $,  $f_1=-f_2=f$, and $g_2=g_1=g$.
Examine first the situation near $T_c$:
\begin{eqnarray}
f &=& \frac{ \Delta }{ \omega^\prime}-\frac{ \Delta ^3}{2 \omega^{\prime\,3}}\,,\quad g  = 1-\frac{ \Delta ^2}{2 \omega^{\prime\,2}}+\frac{3 \Delta ^4}{8 \omega^{\prime\,4}}\, .
\label{fa,ga}
\end{eqnarray}
The self-consistency condition for this situation is
\begin{eqnarray}
\Delta/\lambda =- \pi T\sum_\omega^{\omega_D} f \,.
\label{s-c4}
 \end{eqnarray}
Substituting here $f$ of Eq.\,(\ref{fa,ga}), one has
 \begin{eqnarray}
 \Delta  = -\frac{ \lambda}{2}\left( A\Delta - \frac{D}{2} \Delta ^3\right)\,,
  \label{s-c3}
 \end{eqnarray}
with
 \begin{eqnarray}
A=\sum_0^{\omega_D} \frac{ 2\pi T} { \omega^\prime}&=& \ln\frac{ \omega_D }{2\pi T} -
\psi\left(\frac{\rho+1}{2}\right)\,,\nonumber\\
D=\sum_0^{\infty} \frac{ 2\pi T_c} { \omega_c^{\prime\,3}}&=&-\frac{1}{8\pi^2T_c ^2}\psi^{\prime\prime}\left(\frac{\rho_c+1}{2}\right)\,.
 \label{A,D}
 \end{eqnarray}
 Here,  $\rho_c=1/2\pi T_c\tau$.
Near $T_c$, only terms of the order not smaller than $(\delta t)^{3/2}$ should be  retained. Since $\Delta\propto (\delta t)^{1/2}$,  one can set $T=T_c$   in the coefficient $D$.
Hence,   one obtains:
 \begin{eqnarray}
 \Delta ^2 = \frac{ 2}{D}\left( A + \frac{2}{\lambda}  \right)\,.
  \label{D1}
 \end{eqnarray}
 We now transform the log-term in $A$:
 \begin{eqnarray}
  \ln\frac{ \omega_D }{2\pi T}= \ln\frac{ \omega_D }{2\pi T_{c0}}+ \ln\frac{ T_{c0}}{T_c } +\ln\frac{T_c}{T} \nonumber\\
  = \psi\left(\frac{ 1}{2}\right) +\frac{1}{\tilde\lambda}+\ln\frac{ T_{c0}}{T_c }+\delta t\,,
  \label{log}
 \end{eqnarray}
 where
  the definition of $T_{c0}$, $\ln(2\omega_D e^\gamma/\pi T_{c0})=2/|\lambda|$, has been used.
  Next, we expand the psi-function term in $A$,
 \begin{eqnarray}
   \psi\left(\frac{ \rho+1}{2}\right) =   \psi\left(\frac{ \rho_c+1}{2}\right)  +  \frac{\rho_c}{2}   \psi^\prime\left(\frac{ \rho_c+1}{2}\right)\delta t\,.\qquad
  \label{psi}
 \end{eqnarray}
   Finally, using  Eq.\,(\ref{AG})   for $T_c$, we obtain
\begin{eqnarray}
 \Delta ^2 = -\frac{ 16\pi^2T_c^2}{\psi^{\prime\prime}  }\left(1- \frac{\rho_c}{2}   \psi^\prime\right) \delta t\,,
  \label{D2}
 \end{eqnarray}
where  psi-functions are taken at $(\rho_c+1)/2$.

Now we turn to the functional (\ref{functional}):
\begin{eqnarray}
\frac{ \Omega}{N_n} = -\frac{2\Delta ^2}{\lambda} - 2\pi T  \sum_{\omega }\Big\{  2 [\Delta  f  +\omega(g -1)]-\frac{f ^2}{2\tau} \Big\}  .\qquad
 \label{energy}
 \end{eqnarray}
   Substituting here $f$ of Eq.\,(\ref{fa,ga}) and $\Delta$ of Eq.\,(\ref{D2}) we obtain after straightforward algebra:
\begin{eqnarray}
\frac{ F_s-F_s}{N_n} = \frac{4\pi^2T_c^2}{\psi^{\prime\prime}} \left(2-\frac { \rho_c\psi^{\prime\prime\prime}}{3 \psi^{\prime\prime}}\right) \left(1-  \frac{\rho_c}{2} \psi^{\prime}\right)^2 (\delta t)^2  , \qquad\nonumber\\
 \label{cond-energy}
 \end{eqnarray}
 where the psi-functions are taken at $(\rho_c+1)/2$. The specific heat jump follows:
\begin{eqnarray}
\frac{C_s-C_n}{C_n}\Big |_{T_c} = -\frac{24}{\psi^{\prime\prime}} \left(1-\frac {\rho_c\psi^{\prime\prime\prime}}{6 \psi^{\prime\prime}}\right)\left(1-   \frac{\rho_c \psi^{\prime}}{2} \right)^2  .\qquad
 \label{jump8}
 \end{eqnarray}
 In the clean limit, this gives  $ 12/7\zeta(3)=1.43$.
 Since  $T_c$ can be evaluated for each $\rho_c$, one can plot the jump vs $T_c/T_{c0}$, Fig.\,\ref{f11}(b).

In fact, this behavior of $\Delta C/C_n(T_c)$ is qualitatively similar to the one-band d-wave (although there the clean limit value is 2/3 of 1.43). One can associate this similarity to the fact that in both cases $\langle\Delta\rangle =0$.

\bibliographystyle{apsrev4-1}

\end{document}